# Non-Invasive MGMT Status Prediction in GBM Cancer Using Magnetic Resonance Images (MRI) Radiomics Features: Univariate and Multivariate Machine Learning Radiogenomics Analysis

Ghasem Hajianfar[1,2], Isaac Shiri[2], Hassan Maleki[2], Niki Oveisi[3], Abbass Haghparast[1], Hamid Abdollahi[4], Mehrdad Oveisi[2,5]

1. Department of Medical Physics, Faculty of Medicine, Kermanshah University of Medical Sciences, Kermanshah, Iran
2. Department of Biomedical and Health Informatics, Rajaie Cardiovascular Medical and Research Center, Iran University of Medical Science, Tehran, Iran
3. School of Population and Public Health, The University of British Columbia
4. Department of Radiologic Sciences and Medical Physics, Faculty of Allied Medicine, Kerman University, Kerman, Iran
5. Department of Computer Science, University of British Columbia, Vancouver BC, Canada

**For More Information, Please Contact:**
**Isaac Shiri**
**Department of Biomedical and Health Informatics, Rajaie Cardiovascular Medical and Research Center, Iran University of Medical Science, Tehran, Iran**
**Email: Isaac.sh92@gmail.com**

## Abstract:

***Background and aim***: This study aimed to predict methylation status of the O-6 methyl guanine-DNA methyl transferase (MGMT) gene promoter status by using MRI radiomics features, as well as univariate and multivariate analysis.

***Material and Methods***: Eighty-two patients who had a MGMT methylation status were include in this study. Tumor were manually segmented in the four regions of MR images, a) whole tumor, b) active/enhanced region, c) necrotic regions and d) edema regions (E). About seven thousand radiomics features were extracted for each patient. Feature selection and classifier were used to predict MGMT status through different machine learning algorithms. The area under the curve (AUC) of receiver operating characteristic (ROC) curve was used for model evaluations.

***Results***: Regarding univariate analysis, the Inverse Variance feature from gray level co-occurrence matrix (GLCM) in Whole Tumor segment with 4.5 mm Sigma of Laplacian of Gaussian filter with AUC: 0.71 (p-value: 0.002) was found to be the best predictor. For multivariate analysis, the decision tree classifier with Select from Model feature selector and LOG filter in Edema region had the highest performance (AUC: 0.78), followed by Ada Boost classifier with Select from Model feature selector and LOG filter in Edema region (AUC: 0.74).

***Conclusion:*** This study showed that radiomics using machine learning algorithms is a feasible, noninvasive approach to predict MGMT methylation status in GBM cancer patients

# Introduction:

Glioblastoma multiforms (GBM) are one of the most aggressive malignant brain tumors with an occurrence rate of 2 to 3 cases per 100,000 individuals (1). Post diagnosis, these patients have a median survival time of 15 months, with less than 5% of patients having a 5-year survival time (2). This poor prognosis results from the intra-tumor genetic heterogeneous pattern of GBM (3). Currently, temozolomide (TMZ) is the forefront in therapy for GBM patients. This treatment causes alkylation at the O-6 guanine of DNA, subsequently inducing cytotoxic effects and death in cancer cells (4). Studies show that the methylation status of the O-6 methylguanine-DNA methyltransferase (MGMT) gene promoter could be a predictor for the efficacy of TMZ treatment (5) MGMT is a vital gene which encodes a DNA repair protein (6). Tumor cells with MGMT expression show resistance to TMZ, while tumor cells without MGMT expression are more sensitive to TMZ. Demethylation of the DNA O-6 guanine by MGMT prohibits the DNA degradation induced by TMZ. GBM patients with MGMT promoter silencing show a higher response to TMZ. Several studies have shown that MGMT promoter methylation is associated with longer survival (7-9).

In GBM patients without surgical resection indications, it has been shown that medical imaging could help understand the tumor pathology (10). Clinical studies have indicated that magnetic resonance imaging (MRI) sequences such as T1, T1-contrast, T2, and FLAIR have a feasible role for the prognosis, diagnosis, and treatment plan for GBM (11). Furthermore , further research studies have introduced radiomics as a novel field which can also be used as a powerful prognostic tool (12-14). In combination with MRI sequences, these tools can have a major impact on GBM patient management through higher stratification (15).

Radiomics is an advanced image processing technique which extracts a large number of quantitative features with standard and special algorithms. These features are then correlated with clinical outcomes. The features are shape, intensity, and texture based. The aforementioned features are used during clinical decision making for patient diagnosis, prognosis, and therapy response prediction/assessment (16-18). An extension of this field is radiogenomics, where radiomics features are correlated with genomics parameters (19-21).

Tissue sampling with surgical resection is the gold-standard for the determination of MGMT promoter methylation status. However, this method has limitations such as GBM heterogeneity, a large volume of tissue specimens, and the cost of testing in clinics where it is not routine (22, 23). Recently, radiomics features have been used as imaging biomarkers in MGMT methylation status prediction (24, 25). These quantitative features are reported to simplify the optimum tissue specimen at surgery (26). Multiple studies have explored the efficacy of using these quantitative features in MGMT methylation status prediction. A study by Iliadis *et al*. (27) showed that necrosis volume is inversely associated with MGMT protein–positive tumor cells . Furthermore, a study by Levner *et al* used neural networks as a classifier with S-transform texture features acquired from MRI sequences (T2-w, FLAIR, and T1-w CE) and achieved over 87% accuracy for the prediction of MGMT methylation status (28). Additionally, a study by Eoli *et al* reported that contrast enhanced regions of tumors are correlated with an unmethylated status (29). Moreover, a study by Drabycz *et al* evaluated Visually Accessible Rembrandt Images (VASARI) and automatic texture features for the prediction of MGMT methylation status. Their results revealed that incorporating VASARI with texture features improves the predictive power of MGMT methylation status (30). In another study, Moon et al. (24) found that there are several correlations between MGMT methylation status and features extracted from computed tomography (CT) and T2∗ dynamic susceptibility, contrast enhanced perfusion-weighted imaging, and diffusion tensor imaging MRI features.

This present study aims to predict MGMT status by using T2W and T1W MRI radiomics features, along with univariate and multivariate analysis.

## Material and Methods:

Figure 1 shows the process flow followed in the paper.

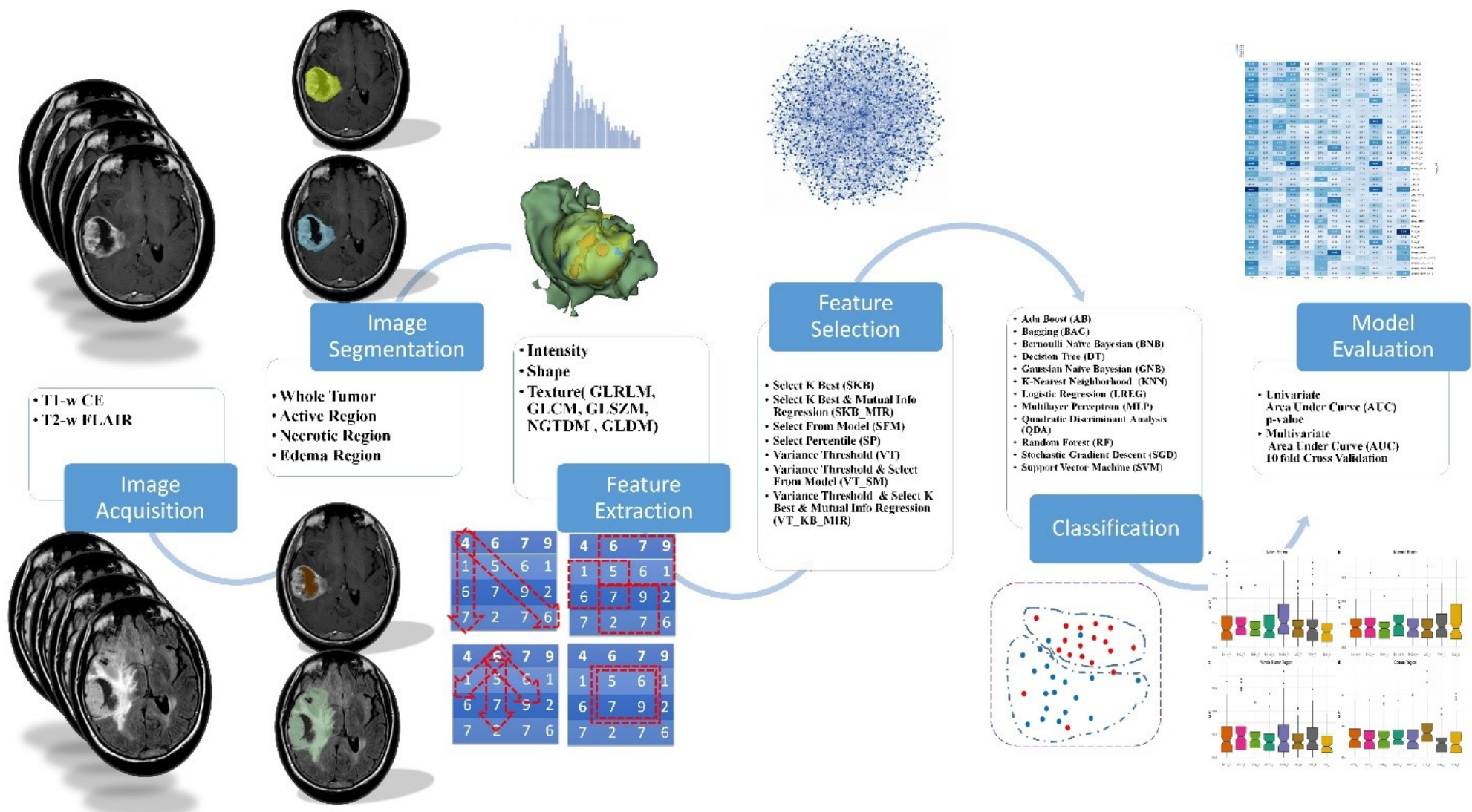

**Figure 1**.Illustrates the process flow followed in the paper.

- **GBM Patient Dataset**

In this study, a total of 122 patients with pathologically confirmed GBM who had two MR images sequences (T2 fluid attenuation inversion recovery (FLAIR) and T1-W CE) were included. All images were downloaded from the cancer imaging archive (TCIA) (31, 32). Clinical data from TCGA (33) and Bady et al. (34) were used to determine the MGMT methylation status using probe cg12434587 and cg12981137 values on HumanMethylation27 and HumanMethylation450 databases. Lastly, 82 patients who had a MGMT methylation status were kept in this study while the remainder of the data was excluded. Table 1 outlines the clinical characteristics of the patients included in this study.

| Characteristic | No.(%) |
|---|---|
| Patient | N=82 |
| Gender | |
| Male | 46 (56.1%) |
| Female | 36 (43.9%) |
| Age, Y | 57.4 ± 13.9 |
| MGMT Status | |
| Methylation | 45 (54.9%) |
| Unmethylation | 37 (45.1%) |
| Karnofsky performance score | |
| <70 | 51 (62.2%) |
| >70 | 23 (28%) |
| Not Available | 8 (9.8%) |

**Table 1.** Clinical characteristics of MGMT status dataset

- **Segmentation**

Tumor were manually segmented with a 3D slicer in the following four regions a) whole tumor (WT), b) active/enhanced tumor (A), c) necrotic regions (N) in T1W CE imaging and d) edema regions (E) in T2 FLAIR imaging. All segmentations were verified by an expert radiologist.

- **Feature extraction**

Before feature extraction, the images were pre-processed. For pre-processing, images were discretized and resampled to 16, 32, 64, 128, 256 gray level bin sizes. Additionally, wavelet (WAV) with different decompositions (HHH, HHL, HLH, HLL, LHH, LHL, LLH and LLL) and Laplacian of Gaussian (LOG) filters with different sigma values (0.5 to 5 with step 0.5) were applied. Feature extraction included three feature sets: shape based, first order, and textures. Thirteen shape features were extracted for each segment along with A/WT ratios, N/WT ratios, and WT/E ratios for the features. Texture sets were Gray Level Co-occurrence Matrix (GLCM), Gray Level Run Length Matrix (GLRLM), Gray Level Dependence Matrix (GLDM), Gray Level Size Zone Matrix (GLSZM), and Neighboring Gray Tone Difference Matrix (NGTDM) (Table 1s in supplemental data). About 7047 features were extracted for each patient, and 41 image sets were prepared for model evaluations (Table 2s in supplemental data).

**Supplemental Table 1.** Radiomics Features

| First Order Statistics (FOS) | Gray Level Co-occurrence Matrix (GLCM) | Gray Level Run Length Matrix (GLRLM) |
|---|---|---|
| - Energy<br>- Total Energy<br>- Entropy<br>- Minimum<br>- 10th percentile<br>- 90th percentile<br>- Maximum<br>- Mean<br>- Median<br>- Interquartile Range<br>- Range<br>- Mean Absolute Deviation (MAD)<br>- Robust Mean Absolute Deviation (rMAD)<br>- Root Mean Squared (RMS)<br>- Standard Deviation<br>- Skewness<br>- Kurtosis<br>- Variance<br>- 19. Uniformity | - Autocorrelation<br>- Joint Average<br>- Cluster Prominence<br>- Cluster Shade<br>- Cluster Tendency<br>- Contrast<br>- Correlation<br>- Difference Average<br>- Difference Entropy<br>- Difference Variance<br>- Joint Energy<br>- Joint Entropy<br>- Informal Measure of Correlation (IMC) 1<br>- Informal Measure of Correlation (IMC) 2<br>- Inverse Difference Moment (IDM)<br>- Inverse Difference Moment Normalized (IDMN)<br>- Inverse Difference (ID)<br>- Inverse Difference Normalized (IDN)<br>- Inverse Variance<br>- Maximum Probability<br>- Sum Average<br>- Sum Entropy<br>- 23. Sum of Squares | - Short Run Emphasis (SRE)<br>- Long Run Emphasis (LRE)<br>- Gray Level Non-Uniformity (GLN)<br>- Gray Level Non-Uniformity Normalized (GLNN)<br>- Run Length Non-Uniformity (RLN)<br>- Run Length Non-Uniformity Normalized (RLNN)<br>- Run Percentage (RP)<br>- Gray Level Variance (GLV)<br>- Run Variance (RV)<br>- Run Entropy (RE)<br>- Low Gray Level Run Emphasis (LGLRE)<br>- High Gray Level Run Emphasis (HGLRE)<br>- Short Run Low Gray Level Emphasis (SRLGLE)<br>- Short Run High Gray Level Emphasis (SRHGLE)<br>- Long Run Low Gray Level Emphasis (LRLGLE)<br>- 16. Long Run High Gray Level Emphasis (LRHGLE) |

| Shape Features | Gray Level Size Zone Matrix (GLSZM) | Gray Level Dependence Matrix (GLDM) |
|---|---|---|
| - Volume<br>- Surface Area<br>- Surface Area to Volume ratio<br>- Sphericity<br>- Compactness 1<br>- Compactness 2<br>- Spherical Disproportion<br>- Maximum 3D diameter<br>- Maximum 2D diameter (Slice)<br>- Maximum 2D diameter (Column)<br>- Maximum 2D diameter (Row)<br>- Major Axis<br>- Minor Axis<br>- Least Axis<br>- Elongation<br>- Flatness<br>- A/T volume ratio<br>- N/T volume ratio<br>- T/E volume ratio | - Small Area Emphasis (SAE)<br>- Large Area Emphasis (LAE)<br>- Gray Level Non-Uniformity (GLN)<br>- Gray Level Non-Uniformity Normalized (GLNN)<br>- Size-Zone Non-Uniformity (SZN)<br>- Size-Zone Non-Uniformity Normalized (SZNN)<br>- Zone Percentage (ZP)<br>- Gray Level Variance (GLV)<br>- Zone Variance (ZV)<br>- Zone Entropy (ZE)<br>- Low Gray Level Zone Emphasis (LGLZE)<br>- High Gray Level Zone Emphasis (HGLZE)<br>- Small Area Low Gray Level Emphasis (SALGLE)<br>- Small Area High Gray Level Emphasis (SAHGLE)<br>- Large Area Low Gray Level Emphasis (LALGLE)<br>- 16. Large Area High Gray Level Emphasis (LAHGLE) | - Small Dependence Emphasis (SDE)<br>- Large Dependence Emphasis (LDE)<br>- Gray Level Non-Uniformity (GLN)<br>- Dependence Non-Uniformity (DN)<br>- Dependence Non-Uniformity Normalized (DNN)<br>- Gray Level Variance (GLV)<br>- Dependence Variance (DV)<br>- Dependence Entropy (DE)<br>- Low Gray Level Emphasis (LGLE)<br>- High Gray Level Emphasis (HGLE)<br>- Small Dependence Low Gray Level Emphasis (SDLGLE)<br>- Small Dependence High Gray Level Emphasis (SDHGLE)<br>- Large Dependence Low Gray Level Emphasis (LDLGLE)<br>- 14. Large Dependence High Gray Level Emphasis (LDHGLE) |

| | | Neighboring Gray Tone Difference Matrix (NGTDM) |
|---|---|---|
| | | - 1-Coarseness<br>- 2-Contrast<br>- 3-Busyness<br>- 4-Complexity<br>- 5- Strength |

**Supplemental Table 2.** Image data sets

| FILTER | SEGMENT | NO. OF FEATURES | NO. OF IMAGE SETS |
|---|---|---|---|
| **SHAPE** | ANTE | 55 | 1 |
| **BIN 16** | A, N, T, E, | 92 | 4 |
| **BIN 32** | A, N, T, E, | 92 | 4 |
| **BIN 64** | A, N, T, E, ANTE | 92 | 4 |
| | SHAPE+ANTE | 368 | 1 |
| | | 423 | 1 |
| **BIN 128** | A, N, T, E, | 92 | 4 |
| **BIN 256** | A, N, T, E, | 92 | 4 |
| **LOG** | A, N, T, E, ANTE | 920 | 4 |
| | SHAPE+ANTE | 3680 | 1 |
| | | 3735 | 1 |
| **WAV** | A, N, T, E, ANTE | 736 | 4 |
| | SHAPE+ANTE | 2944 | 1 |
| | | 2999 | 1 |
| **BIN 64+ LOG+WAV** | A, N, T, E, ANTE | 1748 | 4 |
| | SHAPE+ANTE | 6992 | 1 |
| | | 7047 | 1 |
| **SUM** | | 7047 | 41 |

- **Feature selection**

Seven different feature selections methods were used in the framework and performances were compared (see Table 2).

- **Classifiers**

Twelve Classifiers were implemented and compared (Table 2). The details of each classifier are provided in Table 3s (supplementary).

|   | Feature Selection Methods | Abbreviation |    | Classification Methods | Abbreviation |
|---|---|---|---|---|---|
| 1 | Select K Best | KB | 1 | Adaptive Boost | AB |
| 2 | Select K Best & Mutual Info Regression | KB-MIR | 2 | Bagging | BAG |
| 3 | Select From Model | SM | 3 | Naive Bayes | NB |
| 4 | Select Percentile | SP | 4 | Decision Tree | DT |
| 5 | Variance Threshold | VT | 5 | Gaussian Naive Bayes | GNB |
| 6 | Variance Threshold & Select From Model | VT-SM | 6 | K-Nearest Neighbors | KNN |
| 7 | Variance Threshold & Select K Best & Mutual Info Regression | VT-KB-MIR | 7 | Logistic Regression | LREG |
|   |   |   | 8 | Multilayer Perceptron | MLP |
|   |   |   | 9 | Quadratic Discriminant Analysis | QDA |
|   |   |   | 10 | Random Forest | RF |
|   |   |   | 11 | Stochastic Gradient Descent | SGD |
|   |   |   | 12 | Support Vector Machine | SVM |

Table 2. Feature selection and Classification methods

- **Evaluation**

For univariate analysis, each feature value was normalized to obtain Z-scores, followed by student t-test students for comparison. A p-value of <0.05 was used as a criterion for statistically significant results. Area under the curve (AUC) of the receiver operating characteristic (ROC) curve was used to determine which feature could predict MGMT methylation status. Statistical analysis for this portion was performed in R 3.5.1 (using 'pROC' and 'stats' packages).

For multivariate analysis, an in-house developed python framework was used. A 10 fold cross-validation (CV) was applied for model evaluation. Furthermore, the AUC of ROC curves were also used for model evaluations. Heat-maps and boxplots were constructed to compare different developed models. Lastly, the cross combinations of feature selections and classification methods were depicted as a heat-map (using mean AUC values in cross validation).

## Results:

Table 3 summarizes univariate analyses sorted according to AUC values in top 20 features with different filters and segments. The best predictor of MGMT methylation status was the Inverse Variance feature of GLCM (GLCM_IV) in a Whole Tumor segment with 4.5mm Sigma of Laplacian of Gaussian filter (LOG_4.5S) (AUC: 0.71, p-value: 0.002). Table 4 summarizes the univariate analyses of shape features. Sphericity in the active region (AUC: 0.62, p-value: 0.06) and Elongation and Flatness in edema regions (AUC: 0.62, p-value: 0.08) had better performances in the prediction of MGMT methylation status.

**Table 3.** Top univariate analysis sorted by AUC

| Features | Filter | Region | AUC | p-value |
|---|---|---|---|---|
| GLCM_IV | LOG_4.5S | Whole Tumor | 0.71 | 0.003 |
| GLCM_IV | LOG_5.0S | Whole Tumor | 0.70 | 0.003 |
| GLCM_IV | LOG_4.0S | Whole Tumor | 0.69 | 0.008 |
| NGTDM_Strength | LOG_2.0S | Whole Tumor | 0.68 | 0.050 |
| GLCM_IV | LOG_4.5S | Active | 0.68 | 0.024 |
| FO_Skewness | LOG_4.5S | Edema | 0.68 | 0.010 |
| GLCM_IV | LOG_5.0S | Active | 0.68 | 0.014 |
| GLDM_DV | W_LLL | Necrosis | 0.68 | 0.050 |
| GLSZM_LALGLE | LOG_5.0S | Necrosis | 0.68 | 0.063 |
| NGTDM_Strength | LOG_5.0S | Necrosis | 0.68 | 0.410 |
| GLCM_IV | W_LLL | Necrosis | 0.68 | 0.004 |
| GLDM_LDE | W_LLL | Necrosis | 0.67 | 0.022 |
| GLCM_IDM | W_LLL | Necrosis | 0.67 | 0.010 |
| FO_Kurtosis | LOG_2.5S | Edema | 0.67 | 0.002 |
| NGTDM_Complexity | W_LLL | Necrosis | 0.67 | 0.213 |
| GLCM_Id | W_LLL | Necrosis | 0.67 | 0.012 |
| NGTDM_Strength | LOG_2.0S | Necrosis | 0.67 | 0.055 |
| NGTDM_Strength | LOG_2.5S | Necrosis | 0.67 | 0.273 |
| NGTDM_Complexity | LOG_2.0S | Whole Tumor | 0.67 | 0.148 |
| FO_Kurtosis | LOG_2.0S | Edema | 0.67 | 0.007 |

The Decision Tree classifier with Select from Model feature selector (DT_SFM) in LOG filter in Edema region (E) features had the highest performance (AUC: 0.78). The Ada-Boost classifier with Select from Model feature selector in LOG filter (AB_SFM) in Edema regions (E) features (AUC: 0.74) follows in highest performance (Figure 4s from the supplementary data).

Figure 2 summarizes the results regarding MGMT mutation status prediction based on feature selection and image sets. According to these results, the feature selection performance had a range of 0.52 to 0.61. Additionally, a combination of the Select from Model feature selector (SFM) + the LOG filter in Edema region (E) had the highest performance (AUC: 0.61), followed by a combination of the Variance Threshold and the Select from Model feature selector (VT_SFM) + LOG filter in Edema region (E) (AUC: 0.6).

Figure 3 displays the results regarding MGMT mutation status prediction based on feature classification and image sets. According to these results, classification performance had a range of 0.50 to 0.68. Support Vector Machine classifier (SVM) + combination of Bin 64, LOG and Wavelet (BLW) filters in Necrosis region (N) had the highest performance (AUC: 0.68). This is followed by Ada-Boost classifier (AB) + Bin 16 in Active region (A) (AUC: 0.68), Decision Tree classifier (DT) + Bin 256 in Edema region (E) (AUC: 0.66), and Random Forest classifier (RF) + Bin 256 Edema region (E) (AUC: 0.65).

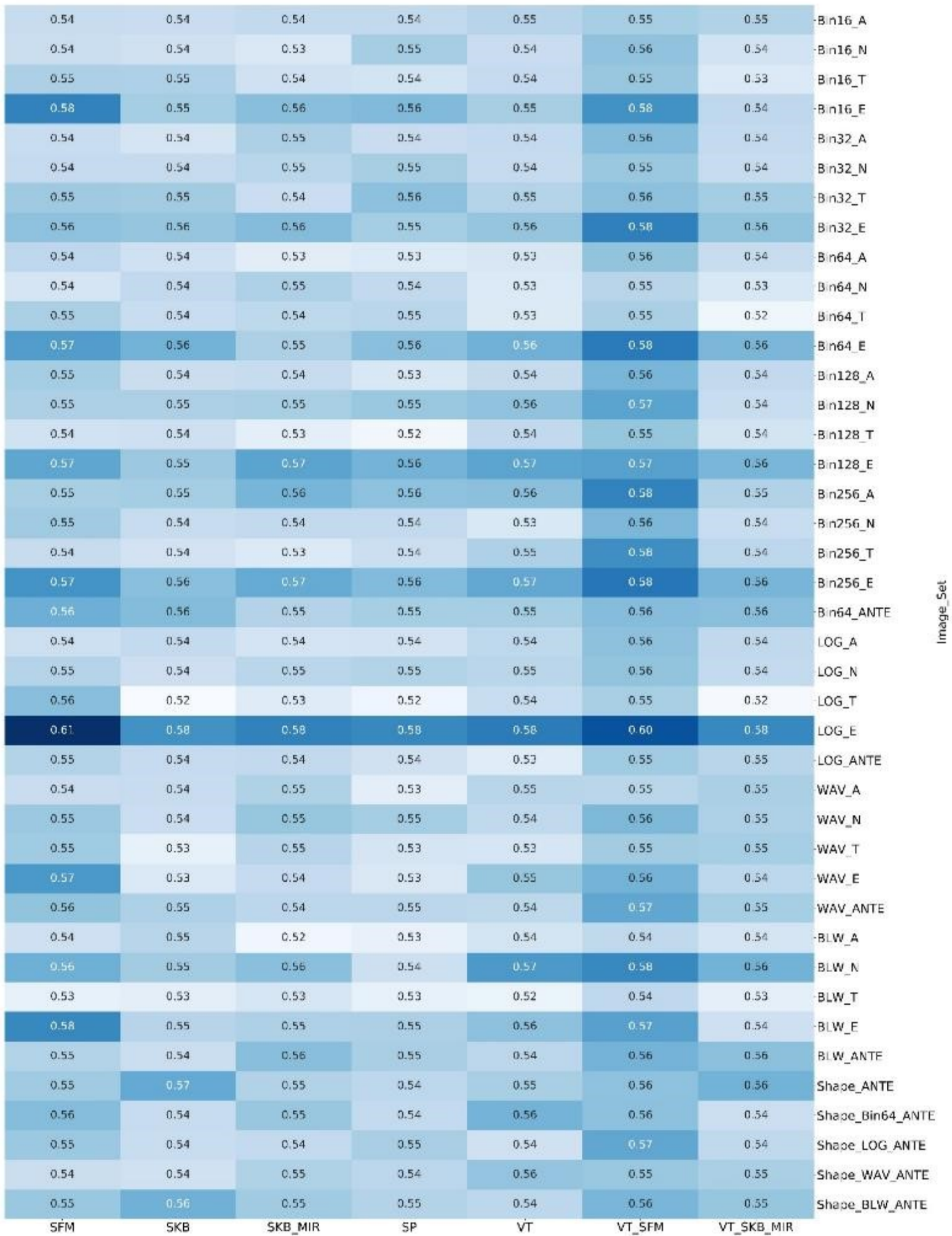

**Figure 2**. Heatmap depicting predictive performance (mean AUC) of image sets (rows) and feature selection methods (columns) in prediction of MGMT methylation status.

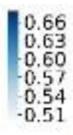
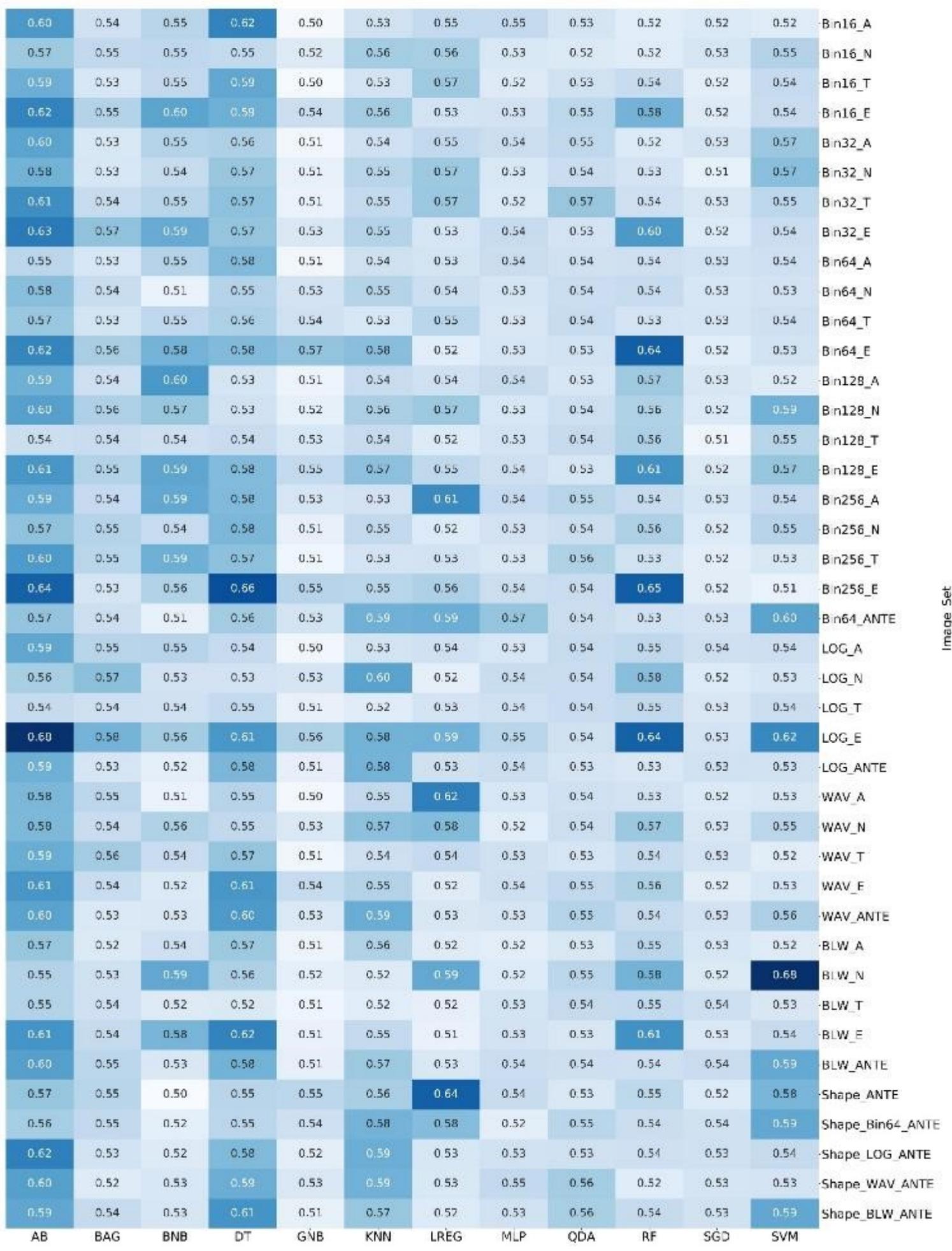

**Figure 3.** Heatmap depicting predictive performance (mean AUC) of image sets (rows) and classifier methods (columns) in prediction of MGMT methylation status.

Figures 4-a and 1s (from the supplementary data) show the results for the prediction of MGMT mutation status in Active tumor regions. It is seen that model performance (mean AUC) had a wide range of 0.50 to 0.68 and Bin 256 had both the highest mean (AUC: 0.55) and the highest predictive performance (AUC: 0.68). Figures 4-b, and 2s (from the supplementary data) show the results for the prediction of MGMT mutation status in Necrosis tumor regions. Notable results include that model performance (mean AUC) had a wide range of 0.50 to 0.68, and a combination of Bin 64, LOG, and Wavelet filters (BLW) images had both the highest mean (AUC: 0.55) and the highest predictive performance (AUC: 0.68).

Figures 4-c and 3s (from the supplementary data) show the results for the prediction of MGMT mutation status in Whole Tumor regions. It should be noted that model performance (mean AUC) had a wide range of 0.50 to 0.69, image set Bin 32 had the highest mean (AUC: 0.55), and image set Bin 256 had the highest predictive performance (AUC: 0.69). Figures 4-d and 4s (from the supplementary data) show the results for the prediction of MGMT mutation status in Edema regions. It is seen that model performance (mean AUC) had a wide range of 0.50 to 0.78, and image set LOG filter had both the highest mean (AUC: 0.57) and the highest predictive performance (AUC: 0.78).

Figures 5-a and 5s (from the supplementary data) show the results for the prediction of MGMT mutation status in a combination of all tumor regions. Results show that model performance (mean AUC) had a wide range from 0.50 to 0.68, image set Bin 64 had the highest mean (AUC: 0.56), and image set LOG filter and combination with shape features had the highest predictive performance (AUC: 0.68). Figure 5-b shows the results regarding MGMT mutation status prediction based on feature selection. A combination of the Variance Threshold and the Select from Model feature selector (VT_SFM) had higher mean predictive performances than other feature selectors (AUC: 0.56). Figure 5-c shows the results regarding MGMT mutation status prediction based on Classifiers. The Ada-Boost classifier (AB) had a higher mean predictive performance than any other Classifier (AUC: 0.59).

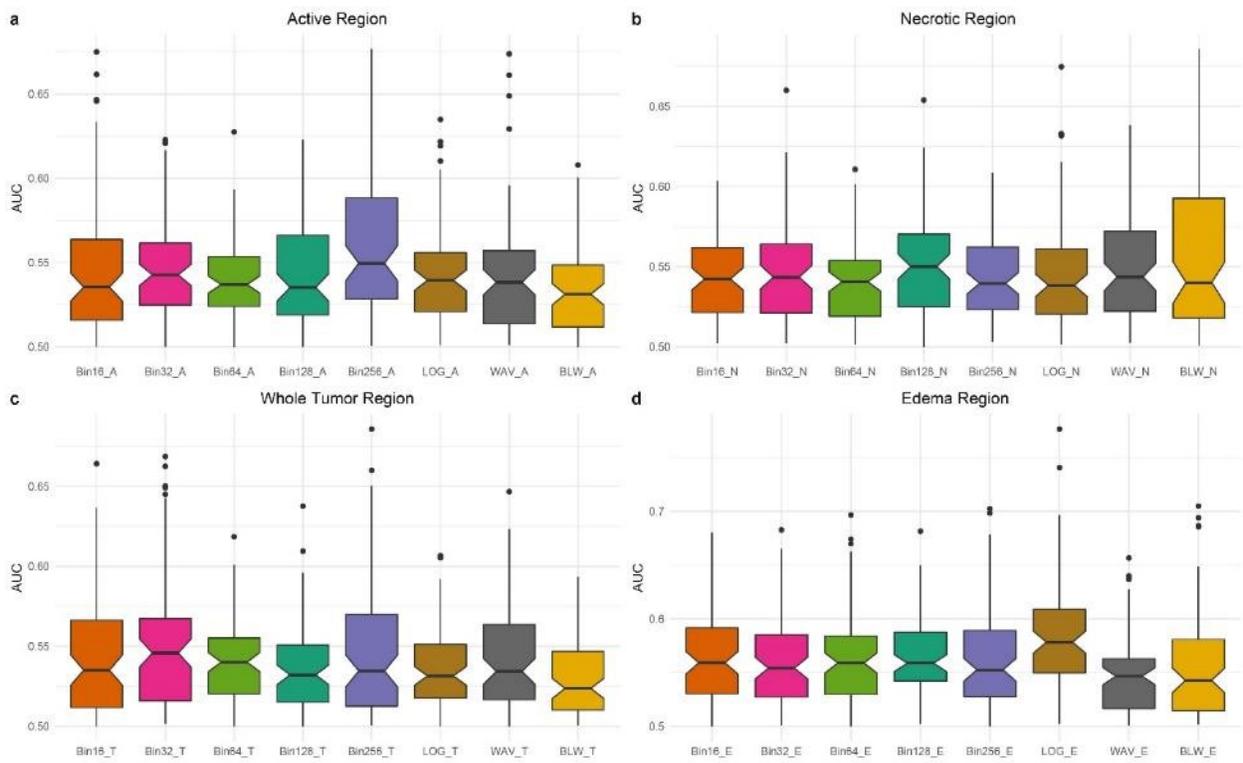

**Figure 4.** (a) Boxplot show predictive performance (AUC) (y-axis) of Active tumor region image sets (x-axis). (b) Boxplot show predictive performance (AUC) (y-axis) of Necrosis tumor region image sets (x-axis). (c) Boxplot show predictive performance (AUC) (y-axis) of Whole

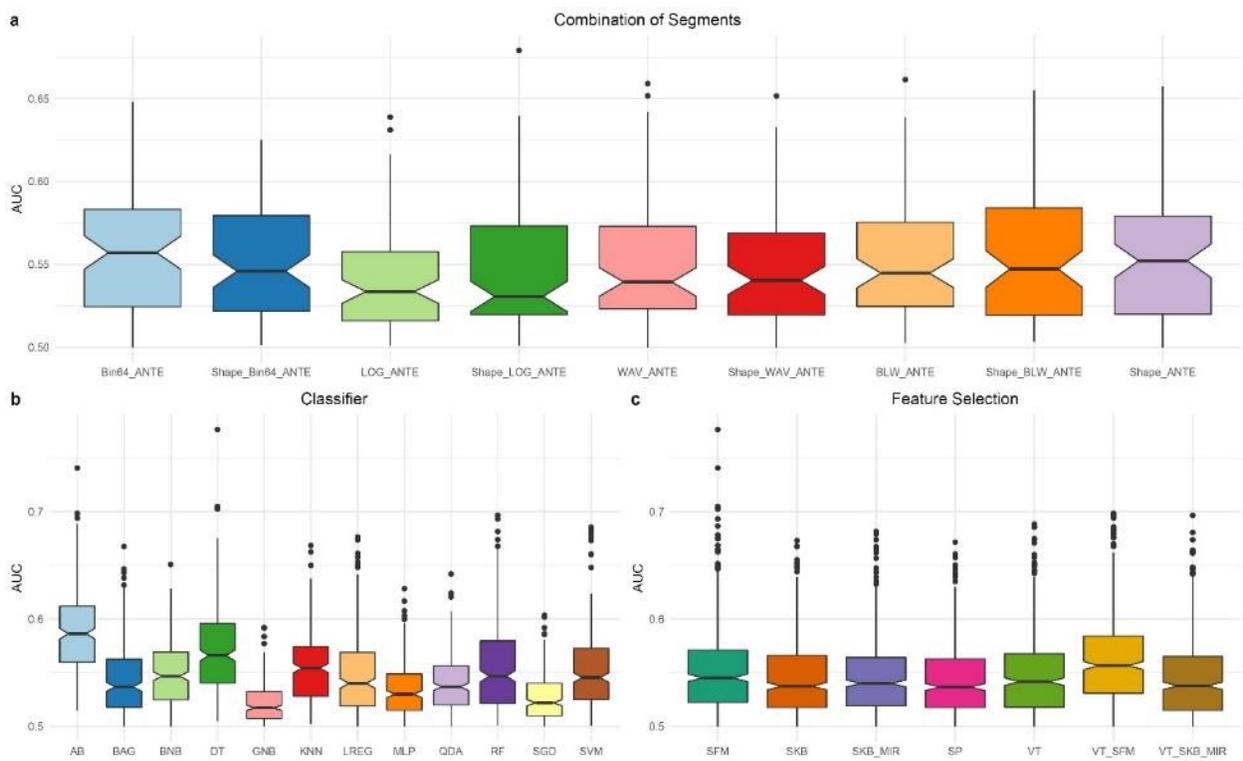

**Figure 5.** (a) Boxplot show predictive performance (AUC) (y-axis) of combination of all tumor regions image sets (x-axis). (b) Boxplot show predictive performance (AUC) (y-axis) of feature selection methods (x-axis). (c) Boxplot show predictive performance (AUC) (y-axis) of classifier methods (x-axis).

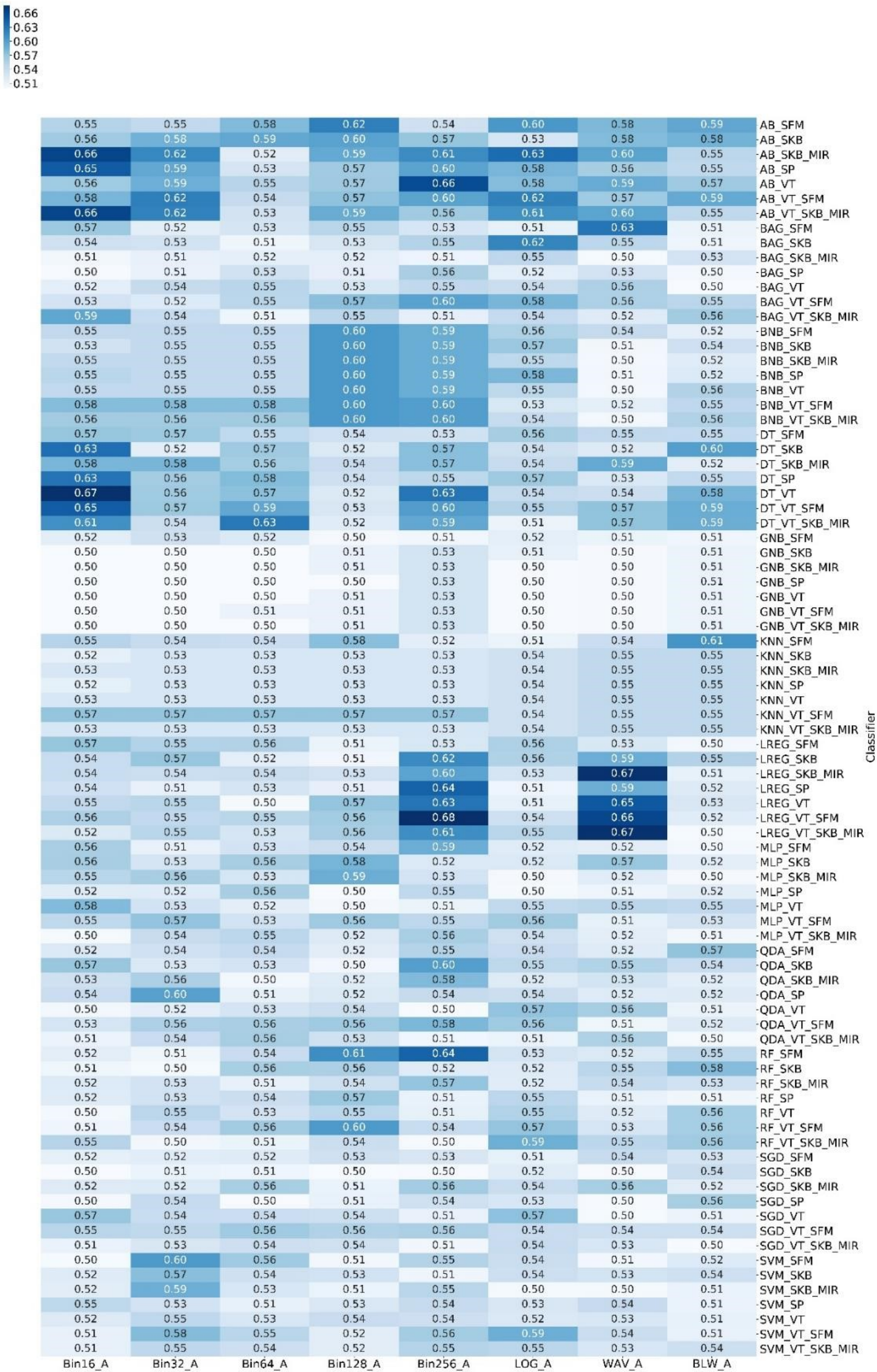

**Supplemental Figure 1.** Heatmap depicting the predictive performance (AUC) of feature selection/classification methods (rows) and Active tumor region image sets (columns) in pre

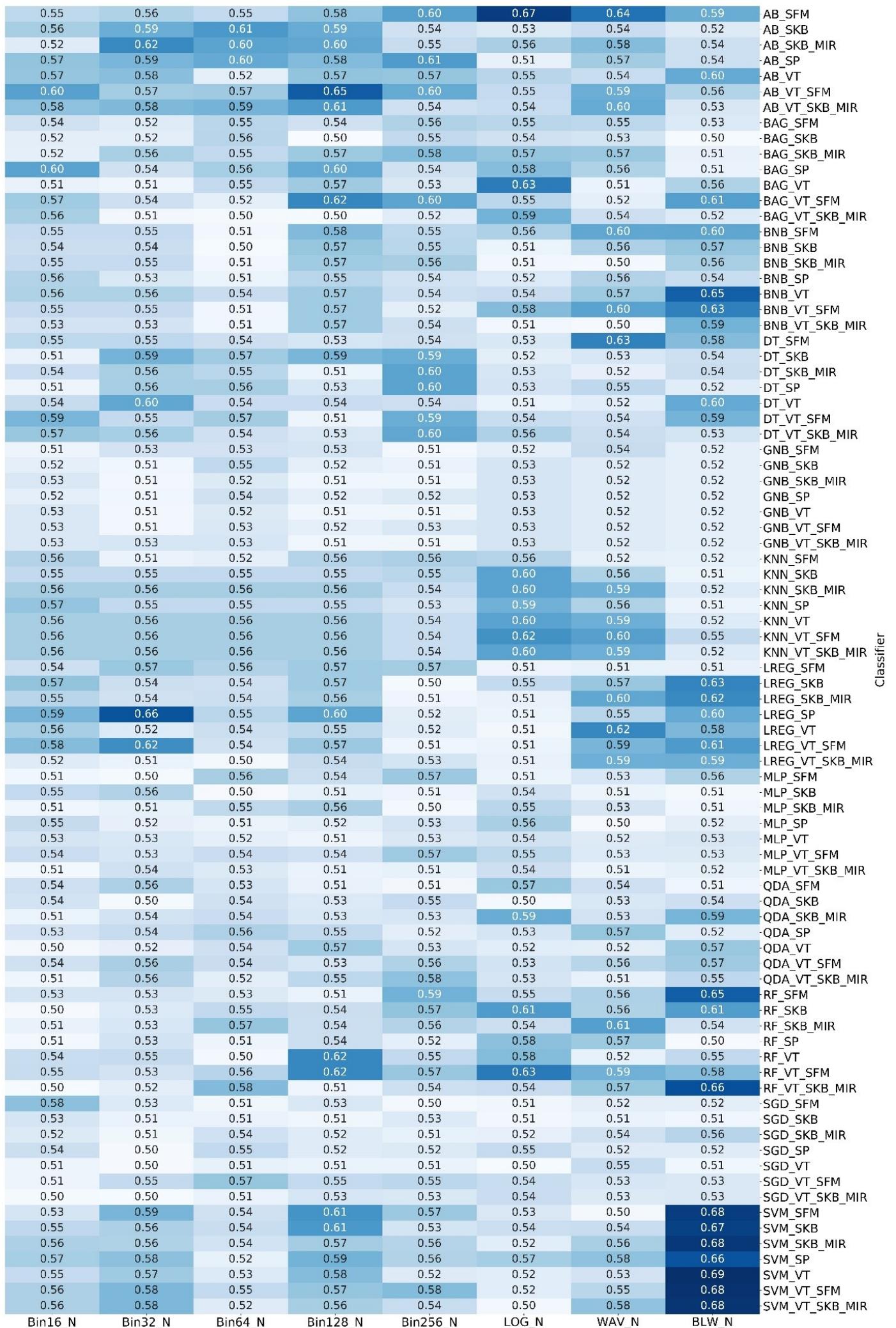

**Supplemental Figure 2.** Heatmap depicting the predictive performance (AUC) of feature selection/classification methods (rows) and Necrosis tumor region image sets (columns) in prediction MGMT methylation status.

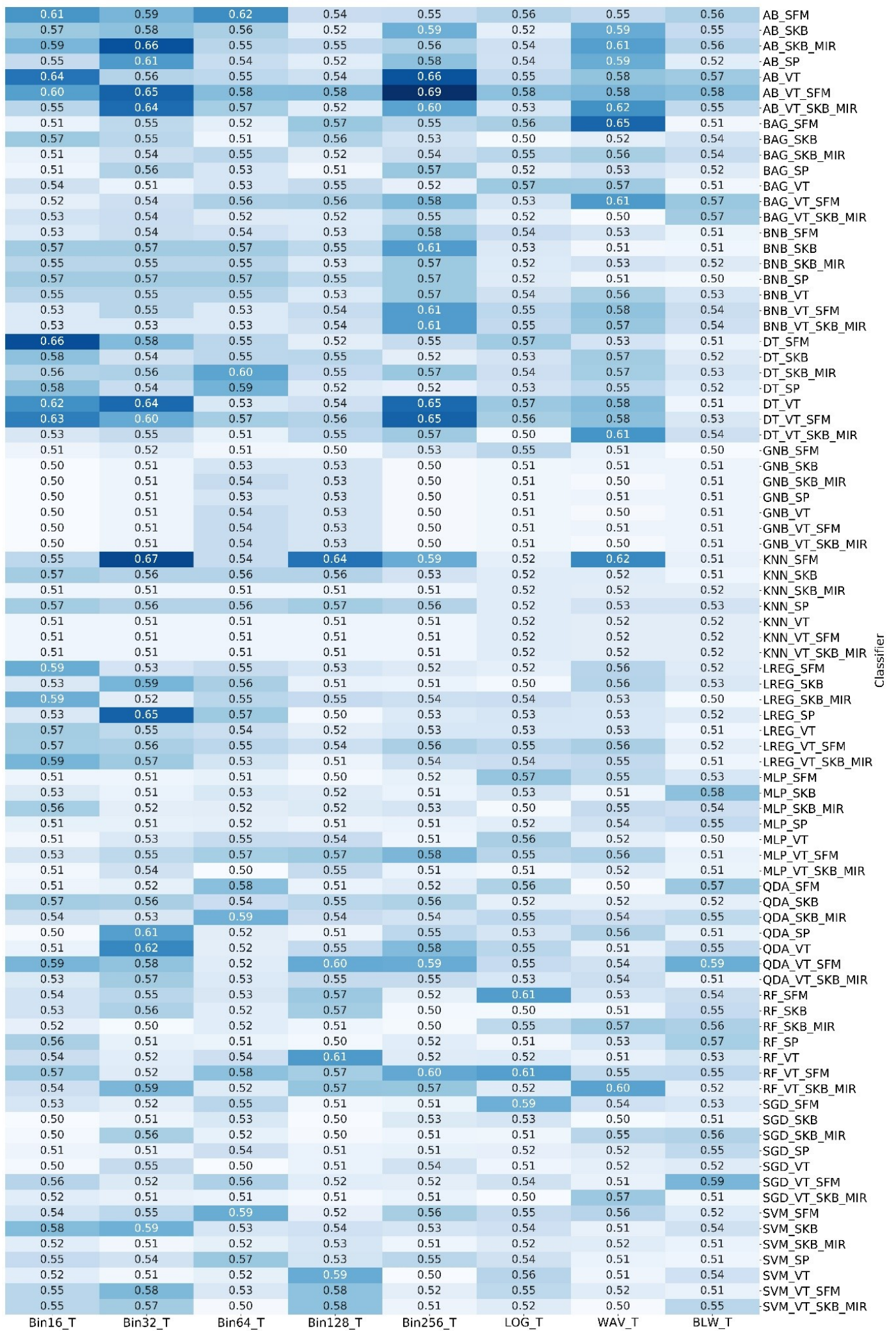

**Supplemental Figure 3.** Heatmap depicting the predictive performance (AUC) of feature selection/classification methods (rows) and Whole Tumor region image sets (columns) in prediction MGMT methylation status.

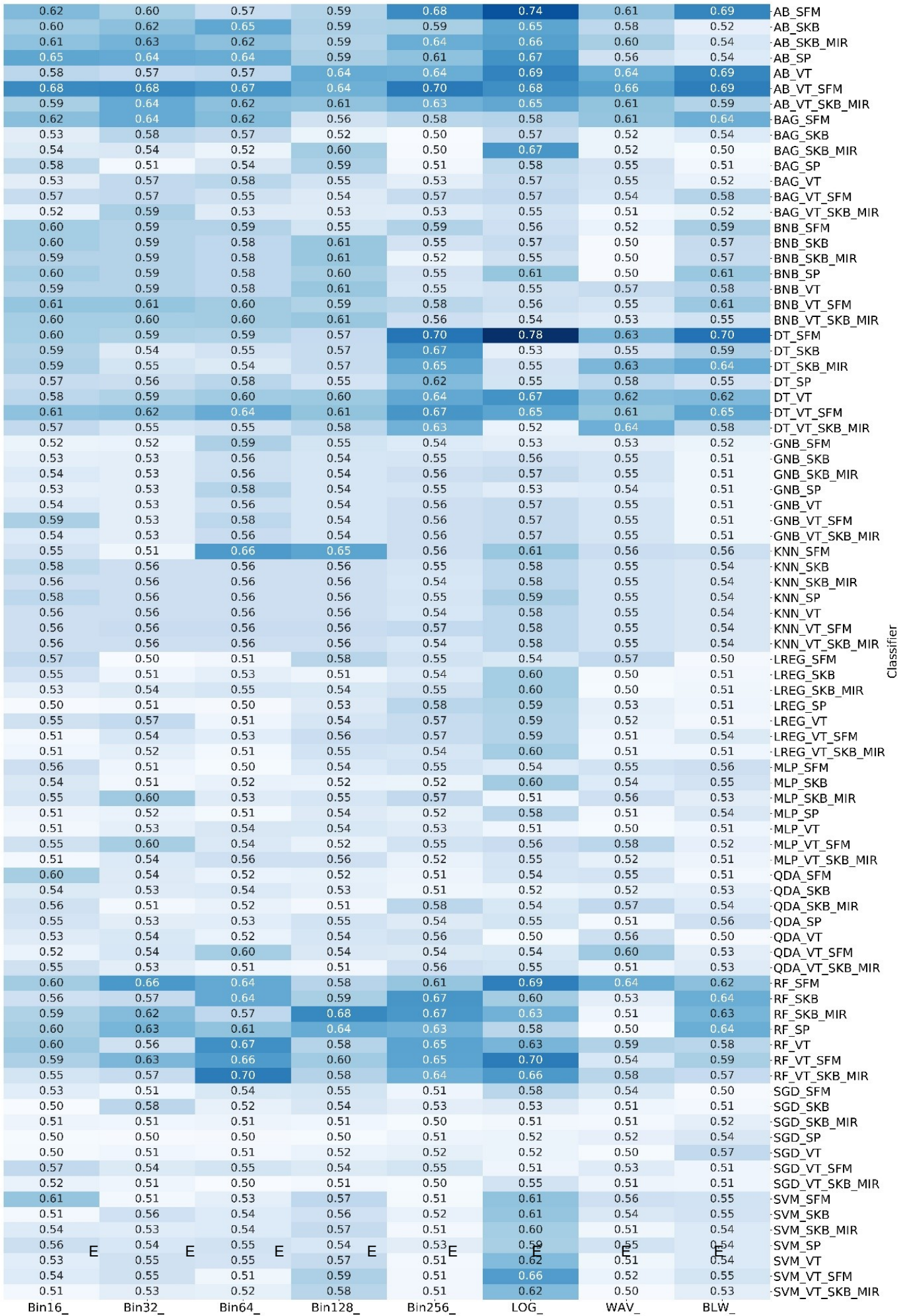

**Supplemental Figure 4.** Heatmap depicting the predictive performance (AUC) of feature selection/classification methods (rows) and Edema region image sets (columns) in prediction MGMT methylation status.

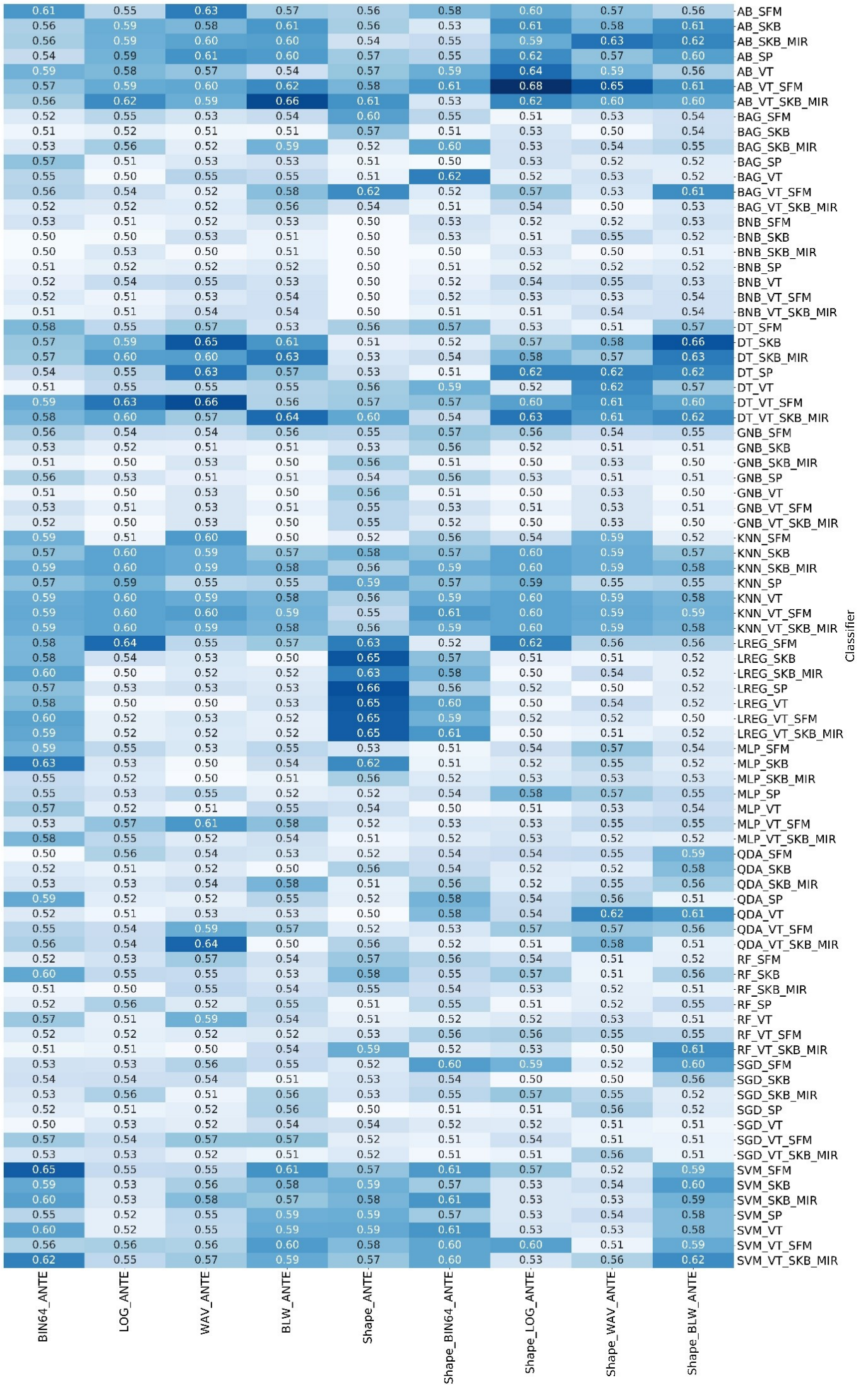

**Supplemental Figure 5.** Heatmap depicting the predictive performance (AUC) of feature selection/classification methods (rows) and combination of all tumor regions image sets (columns) in prediction MGMT methylation status.

**Supplemental Figure 6.** Heatmap depicting the predictive performance (AUC) of feature selection/classification methods (rows) and image sets (columns) in prediction MGMT methylation status

## Discussion:

This study elaborated on an extensive radiomics technique that combines univariate and multivariate analysis with statistical and machine learning methods. This was accomplished by using a large number of features extracted from T1-W CE and T2-FLAIR images taken of 82 GBM patients to predict their MGMT mutation status.

This study showed that GLCM_IV extracted from all segmented regions is an appropriate marker in predicting MGMT methylation status. Also, it was found that the Decision Tree classifier with Select from Model feature selector (DT_SFM) in LOG filter preprocessed in Edema region (E) features had the highest performance (best model, AUC: 0.78). From the classifier and feature selections, AB and VT+SFM had higher mean performances, respectively. In the combinations of 4 segments among 8 image sets, Bin 64 had the highest mean and the LOG filter + shape features image set had the highest predictive performance.

In a study by Li *et al.* (35), they tested the ability of univariate and multivariate analyses in the prediction of MGMT methylation status. Their study developed two multivariate radiomics models to predict MGMT methylation status in GBM cancer patients. They reported that there lacks a statistically significant difference in shape features between methylated and unmethylated MGMT tumors. Several studies have shown that there are no statistically significant differences in shape features between tumors with methylated versus unmethylated MGMT (24, 29, 30, 36, 37). Similar to this current study, the aforementioned studies found features that are correlated with parameters to MGMT methylation and could partake in model prediction. The results from this current study indicate that three shape features (Sphericity in active region, Elongation and Flatness in edema region) have AUC > 0.6 (presented at Table 4). Additionally, figure 5-a shows that combination shape features with LOG filters could improve prediction.

Regarding classification algorithms used to predict MGMT methylation status, Korfiatis *et al.* (26) applied a Support Vector Machine (SVM) and Random Forest (RF) classifier on GLCM and GLRLM features extracted from T1w-CE and T2w images in Enhanced region. Their results showed that the RF classifier had AUC: 0.75 from T1w-CE and AUC: 0.84 from T2w. The SVM classifier had an AUC: 0.76 from T1w-CE and AUC: 0.85 from T2w. This developed model was found to be a high performance predictive model. Performance of our model in the enhanced region is lower than the one found in this study (Figure 1s). However, Korfiatis *et al.* did not extract features from other regions, and the extracted features were limited. Our results show that SVM classifier in combination with all features (BWL imageset) in the necrotic region had the best performance (AUC: 0.68).

Xi *et al.* (25) reported that combinations of sequences improved prediction performance as opposed to singular sequences. They found that T1-WI, T2-WI, and enhanced T1-WI radiomics features could predict MGMT promoter methylation with an accuracy of 86.59% in the training cohort and 80% in the validation cohort. This study extracted features from whole tumors, used SVM for classification, and LASSO for feature selection. Our results show that Bin 64 and LOG filter + shape features with combinations of four segments in T1w-CE and FLAIR has the best performance.

Furthermore, Wei *et al.* (38) reported that ADC values are correlated with MGMT methylation status. However, they also found that T1-W CE and T2 FLAIR sequences had better performances than ADC. Regarding the DWI, the extracted ADC value was found to be a predictor for MGMT promoter methylation status with a sensitivity of 84% and a specificity of 91%. Fusion radiomics signature model had the best performance. Our model performed well in the necrotic region when combining Bin, LOG, and WAV (BLW) filters.

In this study, multiregional segmentation was used for feature extraction in two MR sequences (Figure 1). Regions were also combined to investigate which ones have more prediction power (Figure 5-a). Additionally, the impact of MR image pre-processing and combining regions with differing MLs on the prediction of tumor MGMT mutation status was evaluated. However, this field lacks a comprehensive guideline for the optimal usage of robots features (39-41), classifiers (14). Therefore, this study attempted to determine the best classifier and feature selector for such investigations. Lastly, the impact of pre-processing MR images on MGMT methylation status determination was evaluated. The main limitation of this study was the size of the dataset. To overcome this limitation, models were validated using 10-fold cross-validation to reduce the sensitivity of the results to input data, as well as increase reliability. Future studies should use larger datasets that contain external validation sets.

## Conclusion

This study showed that radiomics using machine learning algorithms is a feasible, noninvasive approach to predict MGMT methylation status in GBM cancer patients. Also, some radiomics could predict the issue alone or in combination with other features. This is important because it could narrow down the treatment options for patients and increase the safety and efficacy of the procedures.